\documentclass[5p]{elsarticle}
\usepackage{amsmath,amssymb}
\usepackage{graphics}
\usepackage{times}

\journal{Physics Letters B}

\begin{document}
\begin{frontmatter}

\title{Upper Energy Limit of Heavy Baryon Chiral Perturbation Theory 
in Neutral Pion Photoproduction}
\author[ucm]{C. Fern\'andez-Ram\'{\i}rez} 
\ead{cefera@gmail.com}
\author[mit]{A. M. Bernstein} 
\address[ucm]{Grupo de F\'{\i}sica Nuclear, 
Departamento de F\'{\i}sica At\'omica, Molecular y Nuclear,
 Facultad de Ciencias F\'{\i}sicas, Universidad Complutense de Madrid, CEI Moncloa,
 Avda. Complutense s/n, E-28040 Madrid, Spain}
\address[mit]{Laboratory for Nuclear Science and Department of Physics,
Massachusetts Institute of Technology,
77 Massachusetts Ave., Cambridge, MA  02139, USA}

\begin{abstract}

With the availability of the new neutral pion photoproduction from the proton  data 
from the A2 and CB-TAPS Collaborations at Mainz it is mandatory to revisit 
Heavy Baryon Chiral Perturbation Theory (HBChPT) and address the extraction of the partial waves
as well as other issues such as the value of the low-energy constants, the energy range 
where the calculation provides a good agreement with the data and the impact of unitarity.
We find that, within the current experimental status, HBChPT with the fitted LECs
gives a good agreement with the existing 
neutral pion photoproduction data up to $\sim$170 MeV and that imposing 
unitarity does not improve this picture. Above this energy the
data call for further improvement in the theory such as the explicit inclusion of the $\Delta$(1232).
We also find that data and multipoles can be well described up to $\sim$185 MeV with Taylor
expansions in the partial waves up to first order in pion energy.

\end{abstract}
\end{frontmatter}

\section{Introduction}  \label{sec:introduction}
Chiral Perturbation Theory (ChPT) is an effective field theory (EFT) 
of Quantum Chromodynamics (QCD) in the low-energy domain where quarks and gluons 
are confined into hadrons and conventional perturbation theory cannot be directly applied. 
Due to the spontaneous breaking of chiral symmetry in QCD
the $\pi$ meson appears as a pseudoscalar Nambu-Goldstone boson \cite{book}
becoming the carrier of the nucleon-nucleon interaction.
However, when fully relativistic spin-1/2 matter fields (i.e. nucleon) are introduced in the theory
the exact one-to-one correspondence between the
loop expansion and the expansion in small momenta and quark masses is spoiled \cite{Gasser88}.
This is due to the fact that the nucleon mass $M$ does not vanish in the chiral limit.
A consistent power counting scheme known as
Heavy Baryon Chiral Perturbation Theory (HBChPT) \cite{CHPT95}
overcomes
this difficulty considering the baryons as heavy (static) 
sources. 
For $\pi N$ scattering and pion photproduction HBChPT has been 
successful at describing experimental data in the near threshold region \cite{CHPT95,CHPT}.  
 In this Letter we address the question of how well it works for the latest and most accurate 
 $\vec{\gamma} p \rightarrow \pi^{0}p$ data to date \cite{Hornidge} and  to provide 
an energy range where HBChPT agrees with
the latest pion photoproduction data, ---
the recently completed Mainz data for  the differential cross sections 
 $d\sigma \slash d\Omega$ and linear polarized photon asymmetries $\Sigma$ 
 for the $\vec{\gamma} p \to \pi^0 p$ reaction taken from threshold through 
 the $\Delta$(1232) region. This was performed with a tagged photon beam 
 with energy bins of 2.4 MeV. We also determined the  
 low-energy constants (LECs) to see if they are actually 
constant as the photon energy is increased. The quality of the HBChPT fits 
--$\chi^2$ per degree of freedom ($\chi^2\slash$dof)-- are also compared to a simple 
 \textit{empirical} benchmark fit, a Taylor expansion of the partial waves.
The data in \cite{Hornidge} are more accurate than previous experiments and the first measurement 
of the energy dependence of $\Sigma$. This has allowed an extraction of the real parts 
of the four dominant multipoles for the first time ---the S-wave $E_{0+}$ and the three P-wave multipoles 
$P_{1,2,3}$ ($E_{1+}$, $M_{1+}$, $M_{1-}$). 
This is a much more significant test of the agreement of HBChPT with experiment. 
As the photon energy increases and the calculations gradually stop 
agreeing with experiment we have determined whether or not this is caused by one particular multipole. 
This information, in addition to the behavior of the low energy constants with photon energy 
provide clues about what improvements are needed to make the HBChPT calculations more accurate. 

\section{Theoretical Framework} \label{sec:theory}
Due to the symmetry breaking,
the S-wave amplitude for the $\gamma p \rightarrow \pi^{0} p$ reaction is small in the 
threshold region, --- vanishing in the chiral limit  \cite{CHPT}. 
Additionally, the P-wave amplitude is large and leads to the $\Delta$(1232) 
resonance at intermediate
energies \cite{AB-Delta}. Hence, for the 
$\gamma p \rightarrow \pi^{0} p$ reaction the S- and P-wave contributions
are comparable  even very close to threshold \cite{AB-fits} and even D waves have 
an important early contribution due to the weakness of the S wave \cite{FBD09}. 
The differential cross section and photon asymmetry 
can be written in terms of electromagnetic responses
\begin{eqnarray}
\frac{d \sigma }{d\Omega}
\left( s, \theta \right) & = & \frac{q}{k_\gamma} W_{T}\left( s, \theta \right) \label{eq:eq1} \\
\Sigma \left( s, \theta \right) &\equiv& \frac{\sigma_\perp - \sigma_\parallel}{\sigma_\perp + \sigma_\parallel} =
-\frac{W_{S}\left( s, \theta \right)}{W_T\left( s, \theta \right) }
\sin^2 \theta \label{eq:eq2}
\end{eqnarray}
where $W_T$ and $W_S$ are the electromagnetic responses, 
$\theta$ is the center of mass scattering angle,
$k_{\gamma}$ the center of mass photon energy, 
$q$ the pion momentum in the center of mass, and
$s$ the squared invariant mass.
The responses  $W_T$ and $W_{S}$
are defined in term of the electromagnetic multipoles:
\begin{equation}
W_{T} =T_0\left( s \right) + T_1\left( s \right) \mathcal{P}_1\left( \theta \right)  
+ T_2\left( s \right) \mathcal{P}_2\left( \theta \right) + \dots \label{eq:wt}
\end{equation}
\begin{equation}
W_{S} = S_0 \left( s \right) +S_1 \left( s \right) \mathcal{P}_1\left(  \theta \right)+ \dots \label{eq:wtt}
\end{equation}
where $P_j \left( \theta \right)$ are the Legendre polynomials in terms 
of $\cos \theta$, the dots stand for negligible corrections, and
\begin{eqnarray}
T_n \left( s \right)&=&\sum_{ij} \text{Re} \{ \: \mathcal{M}^*_i \left( s \right) \: 
T_n^{ij} \: \mathcal{M}_j \left( s \right) \: \} \label{eq:Tn} \\
S_n \left( s \right)&=&\sum_{ij} \text{Re} \{ \: \mathcal{M}^*_i \left( s \right) \: 
S_n^{ij} \: \mathcal{M}_j \left( s \right) \: \} 
\end{eqnarray}
where
$\mathcal{M}_j \left( s \right) =E_{0+}$, $E_{1+}$, $E_{2+}$, 
$E_{2-}$, $M_{1+}$, $M_{1-}$, $M_{2+}$, $M_{2-}$.
The coefficients $T_n^{ij}$ and $S_n^{ij}$ can be found 
in Appendix A in \cite{FBD_PRC09}.

The partial waves (electromagnetic multipoles) are not observables and have 
to be extracted from the experimental data within
a theoretical framework (unless a complete experiment is possible \cite{Bar75}).
In this Letter we employ three approaches to describe S and P waves
that we present in forthcoming paragraphs: 
Section \ref{sec:hbchpt} HBChPT \cite{CHPT96,CHPT01};
Section \ref{sec:uhbchpt}, Unitary HBChPT (U-HBChPT); and 
Section \ref{sec:empirical}, Empirical.
In all cases D waves are incorporated using the customary Born terms.
Higher partial waves can be safely dismissed in this energy region \cite{FBD_PRC09}. 
The conventions employed in this Letter and further information on the
structure of the observables in terms of the electromagnetic multipoles can be found 
in \cite{FBD_PRC09}.

\subsection{HBChPT} \label{sec:hbchpt}
The explicit formulae for the S and P multipoles to one loop
and up to ${\cal O}(q^4)$ can be found in  \cite{CHPT96,CHPT01}.
Due to the order-by-order renormalization process six LECs appear:
$a_1$ and $a_2$ associated with the $E_{0+}$ counter-term:
\begin{equation} 
E_{0+}^{ct}=ea_1 \omega m_{\pi^0}^2+e a_2 \omega^3 \: , \label{e0+ct}
\end{equation}
where $\omega$ is the pion energy in the center-of-mass;
$b_p$ associated with the
$P_3 \equiv 2M_{1+}+M_{1-}$ multipole together with
 $\xi_1$ and $\xi_2$ associated with $P_1\equiv 3E_{1+}+M_{1+}-M_{1-}$ and 
$P_2\equiv 3E_{1+}-M_{1+}+M_{1-}$, respectively.
The $c_4$ LEC associated with $P_1$, $P_2$, and $P_3$
has been taken from \cite{MeissnerNPA00}
where it was determined from pion-nucleon scattering inside
the Mandelstam triangle.
Some other parameters appear in the calculation, but these are fixed.
The full list is: the pion-nucleon coupling constant $g_{\pi N}=13.1$; 
the weak pion decay constant $f_\pi=92.42$ MeV, 
together with the anomalous 
magnetic moments of the proton and neutron,
the nucleon axial charge $g_A$
(which we fix using the Goldberger--Trieman relation $g_A=g_{\pi N}f_\pi/M$); and
the masses of the particles.
The pair  $(a_1, a_2)$ LECs are highly correlated, $r(a_1,a_2)=-0.99$ \cite{FBD09,CHPT01}, 
and it is more convenient to use
the pair of LECs $(a_+=a_1+a_2,a_-=a_1-a_2)$, where $a_+$ 
is the leading order for the counter-term
close to threshold ($\omega \simeq m_{\pi^0}$) \cite{FBD09}.
Henceforth, five LECs are fitted to the data under this approach: 
$a_+$, $a_-$, $\xi_1$, $\xi_2$, and $b_p$.
 
\subsection{U-HBChPT} \label{sec:uhbchpt}
From general principles such as time reversal invariance and unitarity
the S wave can be written as the combination of a 
smooth part and a cusp part \cite{FBD_PRC09,AB-lq,Anant}
\begin{equation}
\begin{split}
E_{0+} =& e^{i\delta_{0}} \left[ A_{0} + 
i \beta q_{+} /m_{\pi^+} \right]  \, ; \, s>s_{thr}^{(\pi^{+}n)} \\
E_{0+} =& e^{i\delta_{0}} \left[ A_{0}  -  
\beta  \left| q_{+} \right| /m_{\pi^+} \right] \, ; \, s<s_{thr}^{(\pi^{+}n)} \: ,
\end{split} \label{eq:FW}
\end{equation}
where $\delta_0$ is the $\pi^0p$ phase shift (which is very small), $\sqrt{s}$ is the invariant mass,
$\sqrt{s_{thr}^{(\pi^{+}n)}}$ the invariant mass at the $\pi^{+}n$ threshold,
$q_{+}$ is the $\pi^{+}$ center-of-mass momentum, $A_{0}$ is $E_{0+}$ in the absence of the 
charge exchange re-scattering (smooth part), and 
$\beta = \text{Re}\left[ E_{0+}\left(\gamma p \to \pi^+ n \right) \right] \times m_{\pi^+} a\left( \pi^+ n \to \pi^0 p \right)$ 
parameterizes the magnitude 
of the unitary cusp and  can be calculated \cite{AB-lq} on the basis of unitarity.
Eq. (\ref{eq:FW})  takes the static isospin breaking (mass differences)  as well as 
$\pi N$ scattering to all orders into account. In the electromagnetic sector it includes 
up to first order in the fine structure constant $\alpha$.
The $\pi^+$ center-of-mass momentum, $q_{\pi^+}$, is real above and
imaginary below the $\pi^+$ threshold; this is a unitary cusp whose magnitude is parametrized by
$\beta$ which can be calculated \cite{AB-lq} on the basis of unitarity 
and taking into account a theoretical evaluation of isospin breaking \cite{Hoferichter}, obtaining 
$\beta=  \left( 3.35 \pm 0.08 \right) \times 10^{-3}/m_{\pi^+}$ where 
$\text{Re} E_{0+}\left(\gamma p \to \pi^+ n \right) 
= \left( 28.06 \pm 0.27 \pm 0.45  \right) \times 10^{-3}/m_{\pi^+}$ \cite{Korkmaz}
and $a\left( \pi^+ n \to \pi^0 p \right)
=\left( 0.1195\pm0.0016 \right) / m_{\pi^+}$ \cite{Baru}.
In HBChPT up to one loop and $\mathcal{O} \left( q^4 \right)$, $\beta$ 
is fixed by the imaginary part of $E_{0+}$ ---that is parameter-free--- 
providing $\beta_{HBChPT} = 2.71 \times 10^{-3}/m_{\pi^+}$ 
which is far away from the unitary value.
Because of the lack of unitarity of the S-wave amplitude \cite{CHPT96}
it is customary to substitute the S wave provided by HBChPT by a 
unitary prescription \cite{FBD_PRC09,CHPT96,CHPT01}.
However, in this Letter instead of substituting the entire S wave for a prescription
we prefer to substitute only the cusp part in $E_{0+}$ from HBChPT 
by the cusp part of $E_{0+}$ in Eq. (\ref{eq:FW}), keeping the smooth part 
provided by HBChPT.
In this way we keep the $E_{0+}$ counter-term and both HBChPT and 
U-HBChPT approaches have the same LECs to fit to the data. 
  
\subsection{Empirical fit} \label{sec:empirical}
The empirical fit is parameterization of the S and P waves with a minimal physics input: 
unitarity in the S wave through the $\beta$ parameter and the angular momentum barrier.
This is accomplished with a Taylor expansion 
in the pion energy in the center of mass $\omega$ 
up to first order on the smooth part of $E_{0+}$  and $P_i\slash q$
adding the cusp part in Eq.  
(\ref{eq:FW}) to the S wave and keeping the imaginary part of the P waves equal to zero, 
in summary\footnote{The empirical parameterization in \cite{Hornidge,FBD_PRC09} 
expands on the photon energy in the laboratory frame $E_{\gamma}$ while 
we prefer to expand in the pion energy in the center of mass frame $\omega$ 
in order to have direct comparison to HBChPT. 
Both approaches render equally good description of the observables 
and provide the same multipoles.}
\begin{eqnarray}
E_{0+} &=& E_{0+}^{(0)} + E_{0+}^{(1)} \frac{\omega-m_{\pi^0}}{m_{\pi^+}}
+ i\beta \frac{q_{\pi^+}}{m_{\pi^+} } \: ,  \label{eq:Swave} \\
P_i \slash q &=&  \frac{P_{i}^{(0)} }{m_{\pi^+}}+  P_i^{(1)} 
\frac{\omega-m_{\pi^0}}{m^2_{\pi^+}}  \: \: \text{;} \: \: i=1,2,3 \label{eq:Pwave}
\end{eqnarray}
where $E_{0+}^{(0)}$, $E_{0+}^{(1)}$, $P_{1}^{(0)}$, $P_{1}^{(1)}$, $P_{2}^{(0)}$, 
$P_{2}^{(1)}$, $P_{3}^{(0)}$, and $P_{3}^{(1)}$ are free parameters 
that will be fitted to the experimental data.
We note that this expansion goes to a lesser order in $\omega$ than HBChPT 
-- i.e. $E^{ct}_{0+}$ in Eq. (\ref{e0+ct}) goes to order $\omega^3$-- 
but entails more parameters.  We note that chiral symmetry is 
not imposed in this approach.

\section{Results} \label{sec:results}
Equipped with the HBChPT, U-HBChPT and empirical approaches 
we perform fits to the experimental data
in  \cite{Hornidge} up to different maximum 
photon energies $E_\gamma^{\text{max}}$ within the range  $\left[ 158.72,191.94 \right]$ MeV 
and compute the $\chi^2\slash$dof as well as the corresponding error bars 
of the extracted parameters (see \ref{sec:errorbars}). 
The energy bins of the data are approximately $2.4$ MeV wide, 
which is taken into account in the fitting and calculations. 
We do not employ the first two energy bins from \cite{Hornidge}, $146.95$ and $149.35$ MeV,
because they are less reliable due to systematic errors, starting the fits at $E_\gamma^\text{min} = 151.68$ MeV. 
The amount of data employed in each fit depends on up to what energy we are fitting,
--- i.e. for our lowest-energy fit ($E_\gamma^{\text{max}}=158.72$ MeV) 
we employ $100$ experimental data 
($80$ differential cross sections and $20$ photon beam asymmetries)
and for our highest-energy fit ($E_\gamma^{\text{max}}=191.94$ MeV) 
we employ $514$ experimental data 
($360$ differential cross sections and $154$ photon beam asymmetries).
The highest-energy fit has been chosen high enough to obtain a $\chi^2\slash$dof 
that ensures that the three approaches no-longer hold and the lowest-energy fit 
to ensure a reliable fit with enough experimental data.
Systematics are not included in the $\chi^2$ 
and this uncertainty can amount up to 4\% in the differential cross section 
and 5\% in the photon asymmetry.
The fits are performed employing a genetic algorithm whose details can be found in \cite{PRC08}.

\subsection{Quality of the fits} \label{sec:chi2}
\begin{figure}
\begin{center}
\rotatebox{-90}{\scalebox{0.34}[0.34]{\includegraphics{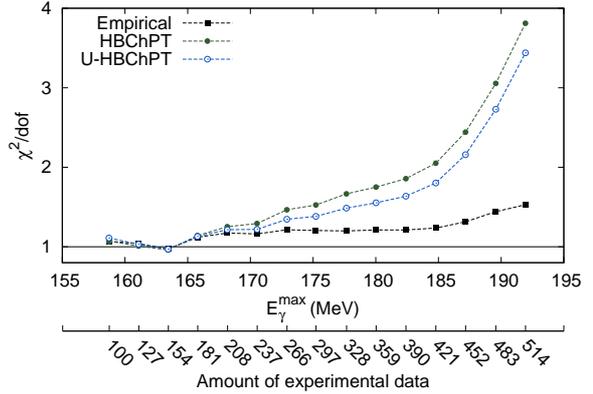}}}
\caption{(Color online.) $\chi^2\slash$dof energy dependence for the
empirical (full black squares), HBChPT (full green circles), 
and U-HBChPT (open blue circles) fits from a minimum photon energy of 151.68 MeV 
up to a variable maximum energy $E_\gamma^{\text{max}}$. 
Each point represents a separate fit and the connecting lines are drawn to guide the eye. 
The points are plotted at the central energy of each bin, although the calculations 
take the energy variation inside of each bin into account. 
The value $\chi^2\slash$dof$=1$ is highlighted with a solid line.} \label{fig:chi2}\end{center}
\end{figure}

Figure \ref{fig:chi2} shows the $\chi^2\slash$dof for every fit performed versus the upper energy 
$E_\gamma^{\text{max}}$ of the fit as well as the number of data.
It is shown that up to $\sim$170 MeV all the fits are equally good providing 
very low $\chi^2\slash$dof.
Above 170 MeV the trend is different; while the empirical fit remains with a good and stable 
$\chi^2\slash$dof, both the HBChPT and the U-HBChPT with the fitted LECs
start rising, a trend that shows clearly 
how the theory fails to reproduce the experimental data above that energy.
Because  we obtain very similar result  for U-HBChPT and HBChPT,
lack of unitarity cannot be blamed for the disagreement between theory and experiment.
The HBChPT result contrasts with the empirical fit that up to 180 MeV 
provides a good description of the data.
Above 185 MeV the $\chi^2\slash$dof of the empirical fit starts to rise 
showing the effects of higher orders 
in the partial waves and the appearance of a non-negligible contribution 
from the imaginary part of the P waves.

\subsection{LECs as a function of $E^{\text{max}}_{\gamma}$} \label{sec:LECs}
\begin{figure}
\begin{center}
\rotatebox{-90}{\scalebox{0.34}[0.34]{\includegraphics{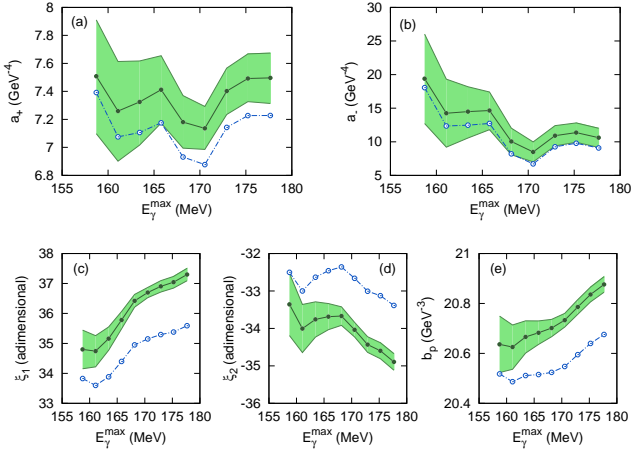}}}
\caption{(Color online.) Upper energy (fit) dependence of the LECs. 
Error band and full circles (green): HBChPT fits; Empty circles (blue): U-HBChPT fit. 
The high correlation between LECs $a_+$ and $a_-$ 
makes their error bars larger. Errorbars for the U-HBChPT LECs are not depicted
but they are approximately of the same size as the HBChPT ones. } \label{fig:LECs}
\end{center}
\end{figure}

An important test of the accuracy of the HBChPT expansion is the stability 
of the empirical LECs versus $E^{\text{max}}_{\gamma}$.
The empirical fit provides a solid
benchmark because the parameters are the same (within errors) in the whole energy region \cite{CD12}.
Figure \ref{fig:LECs} shows the $E^{\text{max}}_{\gamma}$ (fit) dependence of the LECs 
for both the HBChPT (with errors) 
and U-HBChPT approaches.
This includes the S-wave LECs $a_+$ and $a_-$ in 
Figures  \ref{fig:LECs}.(a) and  \ref{fig:LECs}.(b) 
respectively and P-wave LECs
$\xi_1$, $\xi_2$, and $b_p$ in Figures \ref{fig:LECs}.(c),  \ref{fig:LECs}.(d), and  \ref{fig:LECs}.(e).
Errors are larger for the fits with lowest $E_{\gamma}^{\text{max}}$ 
because of the smaller amount of data. 
The S-wave LECs are fairly stable in the whole energy range and both HBChPT and U-HBChPT 
are approximately constant within errors. 
On the contrary, P-wave LECs show a non-stable pattern with a positive slope 
for $\xi_1$ and $b_p$ and a negative slope for $\xi_2$. 
The large error bars make the extracted LECs compatible up to 
$E^{\text{max}}_{\gamma}\sim$175 
MeV except for $\xi_1$, whose value for the 
$E_{\gamma}^{\text{max}}=170.53$ fit is already incompatible with the lower 
energy fit $E_{\gamma}^{\text{max}}=161.08$, confirming that $\sim$170 MeV 
above such energy the theory does not provide a good fit to the data.
Besides, approximately at $\sim$170 the U-HBChPT 
and HBChPT P-wave LECs start to be incompatible.
The U-HBChPT LECs are systematically smaller in absolute value 
than the ones obtained through HBChPT, 
this is expected because the unitary $\beta$ is larger than 
$\beta_{HBChPT}$ giving a larger contribution by the 
$\text{Im}E_{0+}$ which has to be compensated by the other multipoles.
The slopes of the P-wave LECs show that higher order, relativistic 
and $\Delta$(1232) effects are absorbed into them, calling for improvement in the theory.

We have also looked into the correlations  
by computing the correlation coefficient
$r(x,y)= \sigma_{xy} \slash \left(  \sigma_x \sigma_y \right)$
for each pair of parameters and
for every HBChPT and U-HBChPT fit. We find that the correlation
remains more or less stable for each pair throughout every fit.
The S-wave LECs are highly correlated $r (a_+,a_- )  \approx \left[  0.78,0.88 \right]$,
$\xi_2$ and $b_p$ provide
$r  ( \xi_2,b_p ) \approx  \left[ 0.45,0.6 \right]$, which is not unexpected 
due to the photon asymmetry $W_S$ response structure \cite{Hornidge,AB-review},
and the rest are fairly uncorrelated  lying within the range $\left[ -0.2, 0.2 \right]$.
In the case of the S wave the correlation is responsible of 
the large error bars associated to $a_+$ and $a_-$
and indicates that energy dependence and threshold value of $\text{Re}E_{0+}$ 
cannot be obtained separately without further experimental information.
Regarding the magnitude of the
empirical LECs, the empirical values are within a factor of two 
of the values estimated in Refs. \cite{CHPT96,CHPT01} through resonance saturation.

\subsection{Comparison with experimental data} \label{sec:comparison}
\begin{figure}
\begin{center}
\rotatebox{0}{\scalebox{0.45}[0.45]{\includegraphics{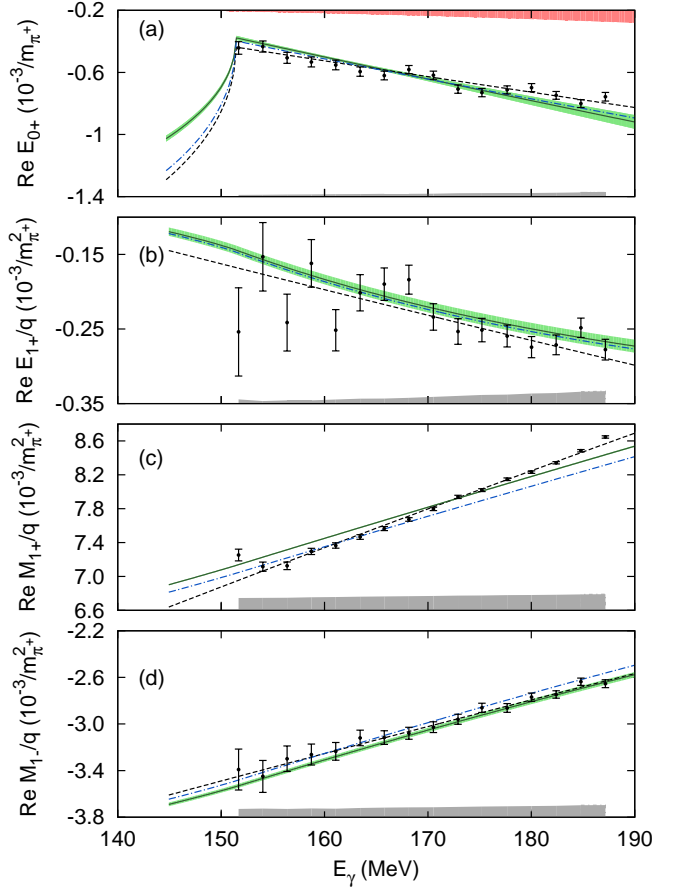}}}
\caption{(Color online.) Real part of the S and P waves. 
HBChPT fit up to 168.16 MeV: Error band and solid line (green); 
Empirical fit up to 180.02 MeV: Dashed.
U-HBChPT fit up to 168.16 MeV: Dash-dotted;
The data shown are the single energy multipoles extracted 
from the experimental differential cross sections and asymmetries 
in \cite{Hornidge}. The gray area above the energy axis represents 
the systematic errors \cite{Hornidge} and the red area at the top of (a)
the uncertainty associated to our knowledge of the D waves.} \label{fig:multipoles}
\end{center}
\end{figure}
In order to compare with experimental data one has to choose a \textit{best fit} 
(set of fitted parameters/LECs) for each approach. 
In our case we pick the fit up to 180.02 MeV for the 
empirical fit ($\chi^2\slash$dof$=1.21$) 
and the fits up to 168.16 MeV for 
HBChPT ($\chi^2\slash$dof$=1.25$) 
and U-HBChPT ($\chi^2\slash$dof$=1.21$) approaches.
Figure \ref{fig:multipoles} shows the single energy 
multipoles extracted from experimental data compared
to the three approaches. The HBChPT fit is shown as an error band.
The procedure to obtain the single energy multipoles from the data is explained in \cite{Hornidge} 
and the error bars are computed as described in \ref{sec:errorbars}. 
The data below the unitary cusp ($146.95$ and $149.35$ MeV) 
are not reliable enough to accurately extract the single 
energy multipoles and, therefore, are not shown.
Overall, the HBChPT and U-HBChPT do a reasonable job describing the multipoles
in the whole energy range (up to $\sim$185 MeV) except in the case of the 
$M_{1+}$, which shows big deviations --specially the slope-- 
between theory and experiment, signaling the necessity to include the $\Delta$(1232) in the analysis.
However, when looking into Figure \ref{fig:multipoles} and comparing fits 
to extracted single-energy multipoles one has to consider that the error bars for both
are computed at the $\chi^2_1$ level as described in \ref{sec:errorbars}
and the impact of systematics (grey band).
Historically the $E_{1+}$ multipole has been considered negligible for many purposes,
an approach that is no longer valid due to the achieved experimental accuracy. 
Moreover, with the current experimental information, 
the inclusion of a non-zero $E_{1+}$ is mandatory to extract accurately 
the two other P-wave multipoles.
Systematically U-HBChPT P waves are smaller in 
absolute magnitude than those extracted through HBChPT. 
This is a consequence of the different $\beta$ value as explained in Section \ref{sec:LECs}. 
The discrepancy at threshold between HBChPT and unitary fits for $\text{Re}E_{0+}$ is also due 
to the value of  $\beta$  \cite{FBD_PRC09}.

\begin{figure}
\begin{center}
\rotatebox{-90}{\scalebox{0.34}[0.34]{\includegraphics{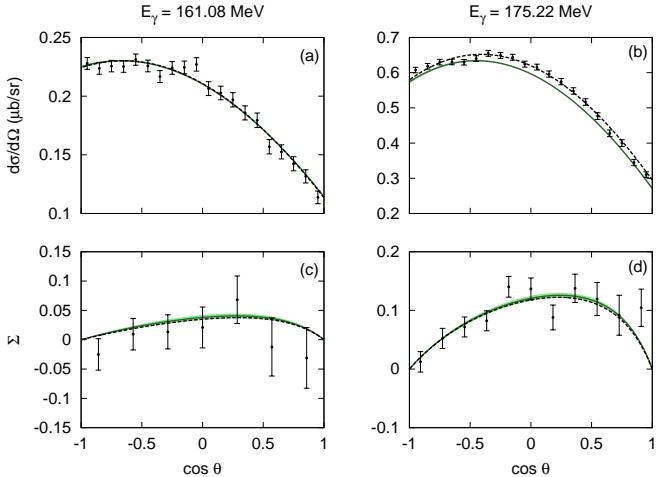}}}
\caption{(Color online.) Differential cross section 
and photon beam asymmetry $\Sigma$.
Error band and solid line (green): 
HBChPT fit up to 168.16 MeV; 
Dashed: empirical fit up to 180.02 MeV.
The U-HBChPT calculation is not depicted because it completely overlaps with the HBChPT.
For the photon asymmetry, (c) and (d), 
the three approaches are undistinguishable.} \label{fig:observables}
\end{center}
\end{figure}
\begin{figure}
\begin{center}
\rotatebox{-90}{\scalebox{0.32}[0.32]{\includegraphics{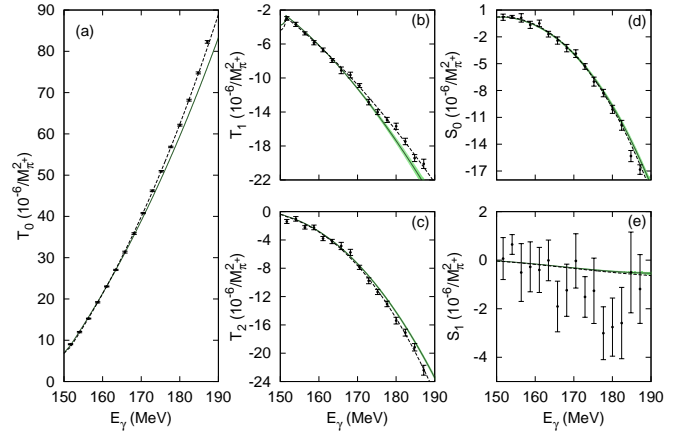}}}
\caption{(Color online.) Components of the $W_{T}$ and $W_{S}$ 
responses for the 
HBChPT fit up to 168.16 MeV (green error band and solid line) 
and empirical fit up to 180.02 MeV (dashed). All the errors are computed at a $\chi^2+1$ level.
If the error bars are increased to $2\sigma$ 
the $S_1$ extraction is compatible with zero in the whole energy range.} \label{fig:Ts}
\end{center}
\end{figure}

Figure \ref{fig:observables} compares the empirical, HBChPT and U-HBChPT approaches to
the differential cross section and photon beam asymmetry at two different energies, one within
the HBChPT and U-HBChPT fitting region ($E_{\gamma}=161.08$ MeV)
and another outside it ($E_{\gamma}=175.22$ MeV).
Two results are noteworthy.
First, the photon asymmetry is well reproduced for both energies by all the fits,
Figures  \ref{fig:observables}.(c) and \ref{fig:observables}.(d); 
if we compare with other energies --higher, lower and intermediate-- 
we find the same level of agreement between theory and data,
obtaining that all the approaches are of the same quality and provide a good description
of the photon asymmetry in the whole energy range considered in this Letter.
Second, the HBChPT and U-HBChPT approaches underestimate the cross section
for energies above the fitting limit ($E^{\text{max}}_{\gamma}=168.16$ MeV),
as can be seen in Figure  \ref{fig:observables}.(b).
This situation is clearer if we look into the component $T_0$ 
of the differential cross section response $W_T$ in Figure \ref{fig:Ts}.(a)
which above $\sim$170 MeV is largely underestimated by the HBChPT approach.
The $T_0$ component is essentially the total cross section and is
dominated by $|M_{1+}|^2$ \cite{FBD_PRC09}, 
which as seen in Figure \ref{fig:multipoles}, is not so well described by the theory.
The significant components of the $W_T$ response $T_0$, $T_1$, and $T_2$
are obtained fitting the differential cross sections for each energy bin 
to Eq. (\ref{eq:eq1}). The same operation is done with the photon asymmetry,
fitting the data to Eq. (\ref{eq:eq2}), extracting $S_0$ and $S_1$.
The other two significant components of the $W_T$ response $T_1$ and $T_2$
are fairly well described up to $\sim$175 MeV. In the case of $W_S$, 
$S_0$ is well determined and HBChPT with the fitted LECs
describes it fairly well in the whole energy range.

\subsection{Probing D waves}
The $S_1$ component of $W_S$ 
in Figure \ref{fig:Ts}.(e) 
is due to the interference among P and D waves \cite{FBD09,FBD_PRC09}.
The empirical values of $S_1$ are consistent with the Born terms contribution of D waves but 
unfortunately the experiment is not accurate enough 
in the photon asymmetry to provide a quantitative measurement. 
The small non-zero effect between 175 and 185 MeV 
disappears if errors are computed at a $2\sigma$ level.
Hence, with the current experimental information 
only four quantities can be accurately obtained from
each energy bin $T_0$, $T_1$, and $T_2$ from the differential cross section and $S_0$
from the photon asymmetry, that allows to extract the four multipoles in 
Figure \ref{fig:multipoles}. If we intend to obtain information on the rest of the multipoles 
we need either more accuracy in the data to
pin down $S_1$ (to obtain information on D waves), 
$T_3$ (P$\times$D interference) or $T_4$ (D waves),
or to measure other observables like the target asymmetry  
(to obtain $\text{Im}E_{0+}$) \cite{AB-review,FandTexperiment}, 
the E asymmetry \cite{FBD_PRC09} (D waves)
or the F asymmetry \cite{FBD_PRC09,AB-review,FandTexperiment} 
($\text{Re}E_{0+}$ and D waves).
With our current knowledge of the P waves, the accurate extraction of $\text{Im}E_{0+}$
and the $\beta$ parameter
from the target asymmetry is feasible. This observable
has been measured at Mainz together with the F asymmetry and
data analysis is currently in progress \cite{FandTexperiment}.

Returning to D waves, they have been incorporated in our analysis as the Born terms
and the $S_1$ component of $W_S$ is consistent with this approach.
Up to order ${\cal O}(q^4)$ in HBChPT this is the only contribution together with 
an $E_{2-}$ counter-term \cite{Hilt2012} which provides an additional LEC. 
However we have neglected it in our calculation
because it has no impact in the $\chi^2$, 
and, therefore, it cannot be determined.
Current experimental information does not allow to test our knowledge 
on D waves but we are hopeful about forthcoming experiments
and we think that
future more accurate data will provide a measure of the D-wave effects 
and allow to pin down the 
$E_{2-}$ counter-term if a deviation from Born terms is found.

\section{Conclusions} \label{sec:conclusions}
Because of the high-quality experimental data gathered by the A2 and CB-TAPS Collaborations
at Mainz we can asses the electromagnetic multipoles and their energy dependence 
to the best precision ever and we can accurately assess the energy range where HBChPT
with the fitted LECs provides a good description of the data.
Based on the accumulated evidence 
--LECs stability, $\chi^2\slash$dof and the empirical fit which works up to $\sim$180 MeV-- 
we find that HBChPT with the fitted LECs provides a good description of the
experimental data up to 170 MeV.
The lack of unitarity in the S wave is not responsible for the disagreement between 
HBChPT and the experimental data as we have proved through the U-HBChPT approach.
The slopes of the P-wave LECs in 
Figure \ref{fig:LECs} show how 
higher order, relativistic and $\Delta$(1232) effects 
are absorbed into them, calling for improvement in the theory.
Some steps have been taking recently to improve the theory, i.e.
Dispersive Chiral Effective Theory \cite{GL}
which combines dispersion relations with ChPT,
and relativistic Chiral Perturbation Theory  \cite{Hilt2012}
which does not provide better agreement with data
than the HBChPT approach \cite{Hornidge}.
We have achieved an unprecedented accuracy in our empirical extraction of the multipoles from the data. 
This has provided a more sensitive test of the HBChPT calculations then has been previously been possible. 
What we have found is that there is a single multipole ($M_{1+}$) that is causing the gradual deviation from experiment (increasing $\chi^{2}$) with increasing energy so this disagreement is probably due to the fact that the $\Delta$(1232) degree of freedom is not being taken into account in a dynamic way \cite{DeltaBCHPT}.

\section*{Acknowledgements}
We thank the A2 and CB-TAPS Collaborations for making available 
the experimental data prior to publication.
C.F.-R. is supported by \textquotedblleft Juan de la Cierva\textquotedblright \:
programme of  Spanish Ministry of Economy and Competitiveness and his
research has been conducted with support by
Spanish Ministry of Economy and Competitiveness 
grant FIS2009-11621-C02-01,
the Moncloa Campus of International Excellence (CEI Moncloa),
and by CPAN,  CSPD-2007-00042 Ingenio2010. 
A.M.B. research is supported in part by the 
US Department of Energy under contract No. 
DE-FC02-94ER40818.

\appendix
\section{Error bar calculation} \label{sec:errorbars}
Error bars have been computed through a Monte Carlo (MC) simulation. 
Once the minimum $\chi^2_{\text{min}}$ has been assessed the $\chi^2_1$ 
defined as $\chi^2_1=\chi^2_{\text{min}}+1$.
Once we have the $\chi^2_1$ we run a MC varying the values of the parameters, 
we compute the corresponding $\chi^2$ for each set of parameter values, 
and we accept those sets which provide $\chi^2\leq\chi^2_1$. 
If enough statistics are collected, 
the boundary of the simulation defines the confidence ellipse 
and the error bars for each parameter \cite{PDG2012}.
We also obtain correlation plots between parameters as well as the correlation matrix.
Once the MC has been run we have a file with thousands of combinations of the parameters 
which are within the $\chi^2_1$ level. We use those sets to compute the bands in the partial waves 
and the observables which are shown in the figures. In this way the error bands in the partial waves 
and the observables take properly into account the correlations among parameters.


\end{document}